\documentclass[aps,prb,preprint,floatfix,twocolumn,showpacs,10pt]{revtex4}
\usepackage{graphicx}

\begin{document}
\title{The Origin of Stokes Shift in Semiconductor Quantum Dots}

\author
{Anjana Bagga$^{\star}$, P. K. Chattopadhyay$^{\dag}$ and Subhasis Ghosh$^{\star} $}

\affiliation{$ ^{\star}$School of Physical Sciences, Jawaharlal Nehru University, New Delhi 110067\\
             $ ^{\dag}$Department of Physics, Maharshi Dayanand University, Rohtak}

\begin{abstract}
The mechanism of Stokes shift in semiconductor quantum dots is investigated by calculating the energy of the excitonic states. We have taken into account all possible contributions to the total electronic energy in the dot, i.e, dielectric mismatch between dot and surrounding medium, the effects of finite barrier height and electron-hole exchange interaction. The Stokes shift is calculated as a function of radius of  dot and compared with experimental data on two different semiconductor based quantum dots. These results provide evidence for exchange splitting of excitonic states, as the mechanism of Stokes shift in quantum dot
\end{abstract}

\pacs{73.22.-f, 73.21.La, 72.80.Ey}
\maketitle

 The red shift of the emission spectra with respect to absorption spectra is known as Stokes shift, which is commonly observed in semiconductor quantum dots(QDs) and the most important  quantity that determines  the optical properties of QDs. However the underlying mechanism of Stokes shift is still controversial. In case of semiconductor QDs, this shift decreases with the increase in radius of the dot and disappears beyond a certain radius.  The red shift, principally, occurs if either the top of the valence band is an optically passive state (say P state) or if the electron and the hole are in a triplet state. Absorption of a photon from the top of the valence band in such cases is not allowed and is possible only from an optically active state lying deeper in the valence band. The exciton, once formed after absorption, can not decay to the top of the valence band by direct dipole transition and hence is called ``dark exciton''\cite{MN95}. Deexcitation eventually takes place with the help of phonons thus giving rise to red shifted photons.  
        
Two different mechanisms\cite{JL00}$^,$\cite{ALE96} have been proposed to explain the origin of Stokes shift in semiconductor QDs, as shown in Fig. 1. Absorption takes place from the S-state which lies deeper than the P-state in the valence band to form an exciton in the singlet state of the electron and hole. According to the first mechanism, (1) of Fig. 1, deexcitation takes place into the P state with the help of phonons giving rise to a Stokes shift by an amount $\Delta E_{SP}$ representing the difference in energy between the exciton states formed with S and P hole states(Ref.2). According to the second mechanism the exciton in the singlet state first thermalizes into a triplet state from where it deexcites with the help of phonons either (a) to the deeper lying original S state, (2) of Fig. 1, giving rise to a Stokes shift by an amount $\Delta E_{ST}$ representing the singlet triplet splitting(Ref. 3), or (b) to the P state at the top of the valence band, (3) of Fig. 1, with the Stokes shift given by $\Delta E_{SP}+\Delta E_{ST}$. In addition to this, the red shift is also observed if the final excited state has a different atomic configuration than the initial ground state. Such a red shift, known as Frank Condon shift, is commonly observed in molecules and point defects in solids and  may be relevant in ultra small QDs\cite{AF03}.

In this Letter we investigate the origin of Stokes shift in III-V(InAs) and II-VI(CdSe) semiconductor based QDs by calculating the energies of excitonic states. We have taken into account the dielectric mismatch effects, the effect of finite barrier height and the effect of the electron hole exchange interaction and then compared the results with experimental data. InAs and CdSe are chosen because of their (i) different crystal structures, (ii) different crystal field effects which is zero in zinc blende structures(i.e InAs) but significantly large in wurtzite structure(CdSe), (iii) different effective masses which is small in InAs, but large for CdSe and (iv) different nature of bonds. It has been observed   experimentally\cite{MN95}$^,$\cite{ALE96}$^,$\cite{UB97} that the red shift decreases with radius and disappears beyond a certain radius. Essentially, the strong dependence of Stokes shift on the radius of the QD, is used to investigate the origin of Stokes shift in QDs based on these semiconductors(InAs, CdSe).  

 Optical transitions from the electron state $\psi_{e}\left(\bf{r}\right)
=j_{0}(\frac{\alpha_{1}^{0}}{R}r)Y_{00}\left(\theta,\phi\right)\left|S\alpha\right\rangle$, where $ \left|S\alpha\right\rangle$ are the Bloch functions of the conduction band, are possible only to hole states in the valence band which have the S-state as a component because 
{\footnotesize $ \int Y_{Lm}Y_{00} d\Omega=\delta_{L,o}\delta_{m,o}$.} The optical transition
  probability from the bottom of the conduction band to a hole state in the valence band is given by \cite{ALE92}
$P_{\alpha\beta}=\left|\int dr  r^2 j_{0}(\frac{\alpha_{1}^{0}}{R}r)
\left(cR_o(r)\right)\right|^{2}
\left|\left\langle u_{\mu}\beta\left|e\widehat{\bf{p}}\right|S\alpha\right\rangle\right|^{2}$,
where c is the component of the S-state $R_{o}(r)$ in the hole wavefunction,
 $\beta$ is the spin projection of the removed electron in the hole state and $u_{\mu}$ are the Bloch wavefunctions of the hole states at the valence band top. $P_{\alpha\beta}$ is zero if $\alpha\neq\beta$. Optical transitions from the bottom of the conduction band to the top of the valence band is not possible if either    
    (1) the hole state does not have a component of the S-state (i.e if c=0), or
    (2) the spin projections of the removed electron in the hole state and the electron in the conduction band are  not equal, $\alpha\neq\beta$. Since the hole spin is obtained by flipping the spin of the removed electron, it follows that optical transitions are not possible if the spins of the electron and the hole are parallel i.e if they are in a triplet state. Thus, if  the top of the valence band is a P-state, or the electron and the hole are in a triplet state, a {\sl dark exciton} is formed.   

The exciton states are obtained from the Hamiltonian     
\begin{equation}
\label{h1}	
H=H_{e}+H_{h}+H_{so}+V_{e-h}+V_{Pol-s}+V_{Pol-eh}	
\end{equation}    
where $H_{e}, H_{h}, H_{so}, V_{e-h}$ are the electronic Hamiltonian, the hole Hamiltonian, the spin-orbit interaction and the e-h coulomb interaction respectively given in Ref.2 and 7. The last two terms represent the surface polarization energies arising due to the difference in the dielectric constants between the semiconductor quantum dot and the surrounding medium\cite{TT93}. $V_{Pol-s}$ is the self energy of the electron and hole due to their image charges and $V_{Pol-eh}$ is the mutual interaction energy between the electron and hole via image charges. The excitonic states obtained from Eq.(\ref{h1}) will be split further due to exchange interaction\cite{ALE96} $H_{ex}$ between the electron and hole.  
        
Since the nature of the state at the top of the valence band plays a pivotal role in the formation of dark and bright excitons, we first investigate the hole states for both the zinc-blende and wurtzite structures using the {\bf k.p} hole Hamiltonian $H_{h}+H_{so}$ given in Ref.9. The energy eigenvalues of the hole states without inclusion of excitonic effects as a function of dot radius for InAs(zinc-blende structures) and CdSe(wurtzite structures) are shown in Figs. 2(a) and 3(a) respectively. These eigenvalues get modified when  coulomb interaction between  electron in the conduction band and  hole in the valence band are taken into account. The corresponding results including excitonic effects are shown in Figs. 2(b) and 3(b) respectively. 

From Fig. 2(a) we observe that in InAs the optically passive $P_{\frac{3}{2}}$ is the ground state which lies below the first optically active state $S_{\frac{3}{2}}$ state. The states are labeled $S_{\frac{3}{2}}$, $P_{\frac{3}{2}}$ etc where the capital letters correspond to the lowest $\bf{L}$ present and the subscripts gives the total angular momentum $\bf{F^{\prime}}$=$\bf{L}+\bf{I}+\bf{S}$, where $\bf{L}$ is the orbital angular momentum, $\bf{I}$(=1) is angular momentum of the Bloch wavefunction at the valence band top and $\bf{S}$ is the spin of the hole. The difference in energy between the optically passive 
$P_{\frac{3}{2}}$ state and the optically active $S_{\frac{3}{2}}$ state decreases as the radius of the dot increases and eventually $S_{\frac{3}{2}}$ becomes the ground state at around $R \sim 50$\mbox{\AA}. Hence dark exciton formation and associated Stokes shift will be observed below this radius.

For the wurtzite structure, the Hamiltonian is axially symmetric and only the z-component 
$M=L_{z}+I_{z}+S_{z}$ of the total angular momentum is conserved. From Fig.3(a) it is observed that for CdSe the optically passive $\left|P_{x}\uparrow\right\rangle$ with $M=\frac{1}{2}$ state is the ground state at $R \sim 10$\mbox{\AA}. The energy of $\left|P_{x}\uparrow\right\rangle$ in units of  $\epsilon_{o}=\frac{\gamma_{1}}{2m_{o}}\left(\frac{\hbar}{R}\right)^{2}$ ($\gamma_{1}$ is a Luttinger parameter) has an upward slope as a function of radius and crosses the relatively flat $\left|S_{x}\uparrow\right\rangle$ state with $M=\frac{3}{2}$ at $R \sim 28$\mbox{\AA}. Dark exciton formation and Stokes shift is predicted for CdSe below $R \sim 28$\mbox{\AA}. The trend of S-state and P state can be explained by observing that $\left|S_{x}\uparrow\right\rangle$  with $M=\frac{3}{2}$ is predominantly made from only one state with composition {\footnotesize $L=0, I_{z}=+1, S_{z}=+\frac{1}{2}$}
whose energy decreases as $\frac{1}{R^{2}}$ but in units of $\epsilon_{o}$ is independent of R and hence the state is relatively flat. But the optically passive P state with $M=\frac{1}{2}$ can be made either from the state with composition {\footnotesize$L_{z}=-1, I_{z}=+1, S_{z}=+\frac{1}{2} (J_{z}=I_{z}+S_{z}=+\frac{3}{2}) $} or the state with the composition {\footnotesize$L_{z}=+1, I_{z}=-1, S_{z}=+\frac{1}{2} (J_{z}=-\frac{1}{2})$}. These two states are separated from each other by spin-orbit interaction. At very low radii spin orbit effects are negligible relative to kinetic energy terms due to confinement. Hence the two states are almost degenerate at $R \sim 10$\mbox{\AA}. The coupling between these two states give rise to a splitting of the states with the lower P state lying below the S state. As the radius increases the energies of both the S and P states decrease. The energy of S state decreases as $\frac{1}{R^{2}}$ but the energy of the P state decreases more slowly
because of the presence of the spin-orbit interaction energy which becomes important as the kinetic energy starts decreasing with the increase of radius. Eventually the two levels cross. In Figs. 2 and 3 the energies have been shown in the units of $\epsilon_{o}$.

When the coulomb interaction between the electron in $1S_{e}$ state and the hole is taken into account, the energies decrease. The decrease in energy of S-state is more than that of  P state, the overlap between  $1S_{e}$ state and  S state of the hole being greater. This causes the crossing between S and P states to occur at lower values of R as can be seen from comparison of Figs. 2(b) and 3(b) with the corresponding Figs. 2(a) and 3(a). The differences between the excitonic S and P states $ \Delta E_{SP}$ as a function of R, are shown in the insets to Figs. 4 and 5. 
       
The surface polarization charges arising due to the dielectric mismatch between the dot and the surroundings also affect the excitonic energies. The contributions of self energy $V_{Pol-s}$ and attractive energy $V_{Pol-eh}$ are opposite in sign and comparable in magnitude\cite{VAF02}. Hence the dielectric mismatch effects are greatly reduced with the net effect being an increase in energy  depending on the amount of dielectric mismatch between the dot and the surrounding.  The increase in energy for the P state is more than that for the S state resulting in a crossover to occur at a lower dot radius. This can be seen from Figs.2(b) and 3(b) which gives the comparison between zero and finite dielectric mismatch, i.e  $\epsilon=1$ and $\epsilon=3$. The self energy of a charge confined to a spherical dielectric is minimum\cite{LEB83} at r=0. Since $\left\langle r^{2}\right\rangle$ for the P-state is larger compared to the S-state, the self energy   $V_{Pol-s}$ is higher for the P-state. But the attractive energy $V_{Pol-eh}$ is higher for the S state compared to the P state. This leads to a higher net increase of polarization contribution $\left[V_{Pol-s} + V_{Pol-eh}\right]$  for the P-state as compared to the S-state and hence to a shifting of the S-P crossover to lower values of R as can be seen  from Figs. 2(b)and 3(b).

Another cause of dark exciton formation and the consequent Stokes shift is the splitting of the active exciton states due to electron hole exchange interaction\cite{ALE96} 
$ H_{ex}= \left(-\frac{2}{3}\right) \epsilon_{ex}\left(a_{o}\right)^{3}\delta\left(\vec{r_{e}-\vec{r_{h}}}\right)\bf{\sigma}.\bf{J}$ 
where $\epsilon_{ex}$ is the exchange strength constant, $a_{o}$ the lattice constant,
$\bf{\sigma}$ the Pauli spin matrices representing the electron spin and $\bf{J}$
is the hole spin matrix.
 The exchange interaction gives rise to a splitting between triplet and singlet state of the electron and hole, with the triplet state lying lower in energy. The triplet state being passive, excitation takes place to the higher singlet state. Deexcitation can take place from the triplet state reached by thermalization, thus giving rise to a red shift of the emmision spectra with respect to the absorption spectra.

In InAs the first optically active S state (Fig. 2(a)) has total anguar momentum 
$ \bf{F^{\prime}}$ $ =\frac{3}{2}$($\bf{L}$  = 0, $\bf{I}$=1, $\bf{S}$=$\frac{1}{2}$)    
and is fourfold degenerate. The $1S_{e}$ electron state in the conduction band being doubly degenerate, the exciton states are eightfold degenerate in the absence of exchange interaction. The exchange interaction splits these degenerate levels into two: the lower optically passive triplet state with $ \bf{F}$=2($\bf{F}=\bf{F^{\prime}}+\bf{S_{e}}$) and the upper optically active singlet state with $ \bf{F}$=1 (Fig. 2(c)). The difference in  energies $\Delta E_{ST}$ between the singlet and triplet states as a function of dot radius is shown in Fig. 4. $\Delta E_{ST}$ decreases with R and vanishes at $R \sim 35$\mbox{\AA}. The exchange interaction is enhanced with decrease in radius because of the increase in the overlap of electron and hole wavefunctions due to confinement.
        
In CdSe (Fig. 3(c)) the ground state with $F_{z}=\pm2$ can be obtained  by coupling a hole with $M=\pm\frac{3}{2}$ with the electron state with  $\left(S_{e}\right)_{z}=\pm\frac{1}{2}$. Since these combinations form triplet states, the ground state with  $F_{z}=\pm2$ is optically passive and forms a dark exciton. The first excited state with $F_{z}$=$\pm 1^{L}$ can be obtained either by coupling $M$=$\pm\frac{3}{2}$, $(S_{e})_{z}$=$\mp\frac{1}{2}$  or alternatively by $M$=$\pm\frac{1}{2}$ with $(S_{e})_{z}$=$\pm\frac{1}{2}$. The state $\pm1^{L}$ will be a mixture of singlet and triplet states and will be optically active. The difference in  energies $\Delta E_{ST}$ between the optically passive ground state and the first optically active  excited state as function of R for CdSe is shown in Fig. 5. As in the case of InAs, $\Delta E_{ST}$ for CdSe also decreases with R and vanishes around $R \sim 50$\mbox{\AA}.

As seen from Figs. 2, 3, 4 and 5 the curves corresponding to finite barrier case lie lower than infinite barrier case. Both the exchange interaction and coulomb interaction depend on the amount of overlap of electron and hole wavefunctions, which is reduced in the finite barrier case due to penetration of the substantial amount of electron wavefunction outside the dot resulting in the decrease of exchange energy  and coulomb binding energy.

Having theoretically calculated $\Delta E_{SP}$ and $\Delta E_{ST}$ we are in a position to  
investigate the various mechanisms for Stokes shift given in Fig. 1 by comparison with experimental data. In Figs. 4 and 5 are shown the plots of $\Delta E_{ST}$  as a function of the dot radius together with the experimentally observed Stokes shift for CdSe and InAs respectively. $\Delta E_{SP}$ as a function of R for these cases are shown in the insets of the corresponding figures. It is observed that $\Delta E_{SP}$ is  much larger than the experimental red shift for InAs. For CdSe, as seen from the inset of Fig. 5 at $R \sim 10$\mbox{\AA} $\Delta E_{SP}$ is almost the same as the experimental Stokes shift but for larger values of R the experimental values are much higher. It is seen from Figs. 4 and 5 that $\Delta E_{ST}$ fits the experimental data rather well. We observed that the fit for CdSe corresponds to an infinite barrier at the dot radius but for InAs a good fit is obtained for a finite barrier height. This can be explained by observing that the effective mass of electron in InAs is very small and hence the lowest eigenvalue lies near the top of the barrier height and there is substantial probability of finding the electron outside the dot. For CdSe, the effective mass of electron is large and the lowest eigenvalue lies near the bottom of the potential well, making the infinite barrier effects predominant. In Fig. 5 we have also plotted the theoretical curve for Stokes shift obtained by Efros et al\cite{ALE96} in which the coulomb interaction between the electron and hole was not taken into account. We observe that better agreement with experimental Stokes shift is obtained by our theoretical curve which takes into account the coulomb interaction.

From a comparision of the theoretical results with experimental data we conclude that the Stokes shift observed in QDs is caused by the splitting of the exciton states by the electron hole exchange interaction through the deexcitation mechanism (2) in Fig. 1.

\section*{Figure Captions}
\noindent

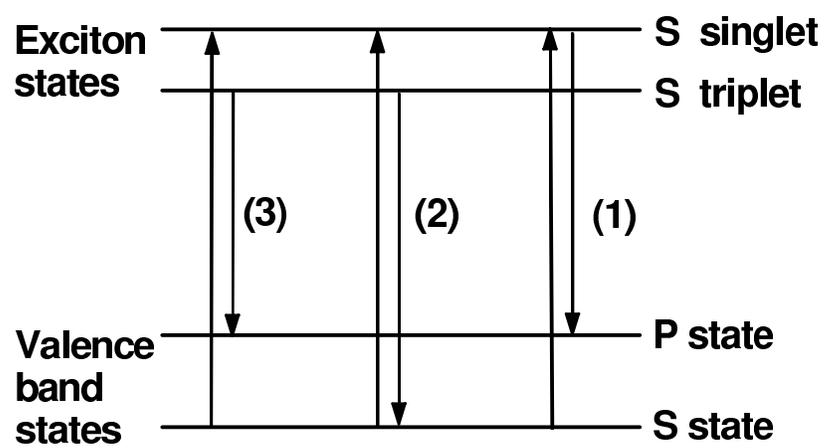
\noindent Fig.1:
The schematic representation of excitation-deexcitation processes taking place at the band edges. 

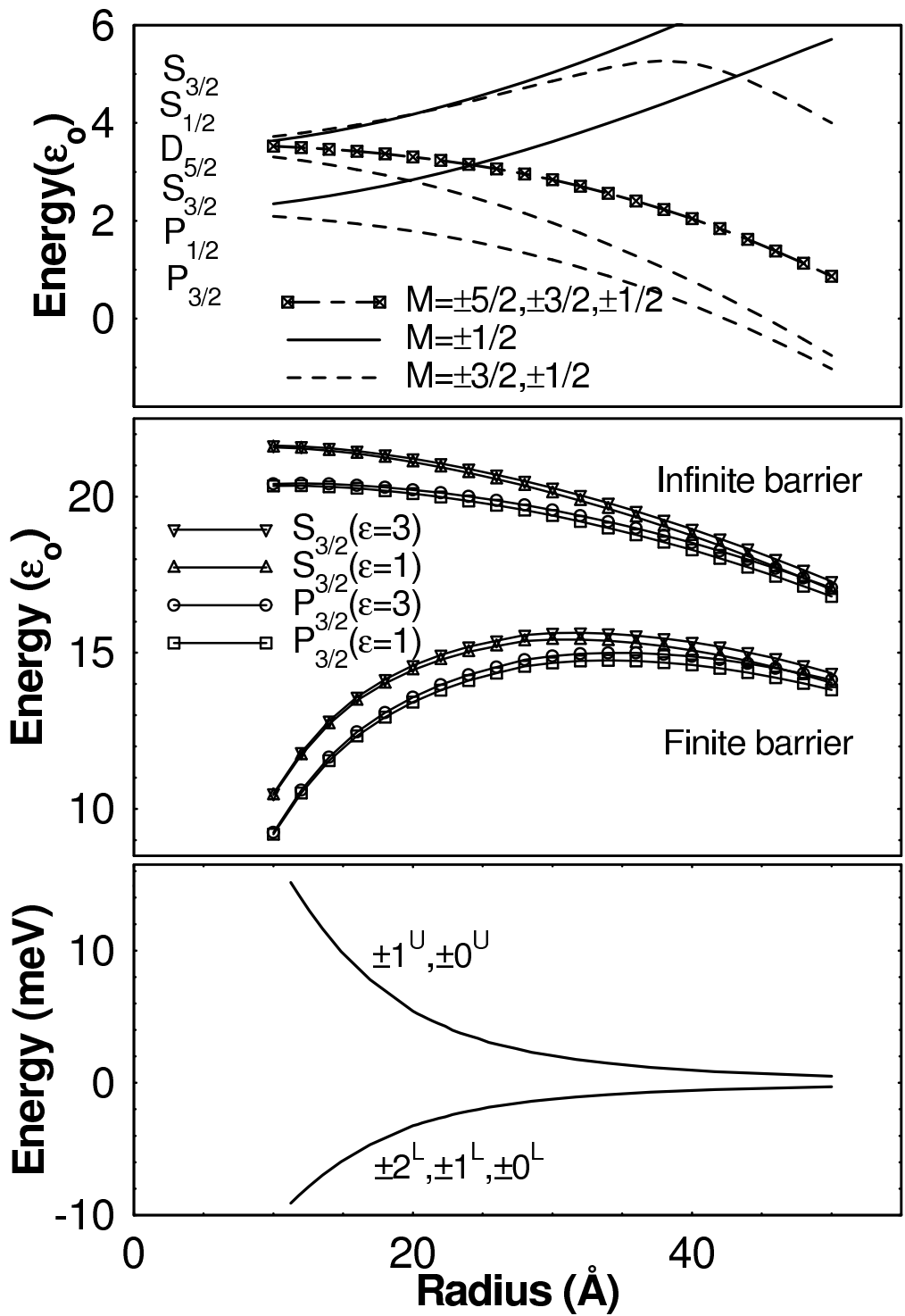
\noindent Fig.2:
The energy levels in units of $\epsilon_{0}$=$\frac{\gamma_{1}}{2m_{o}}\left(\frac{\hbar}{R}\right)^{2}$ for zinc blende InAs QDs as a function of R in the (a) absence and (b) presence of e-h coulomb interaction(without exchange). The energy reference is taken as the top of the valence band without the spin orbit effects included. Eigenvalues in the absence($\epsilon=1$) as well as presence($\epsilon=3$ of dielectric mismatch effects are also shown in (b) for both the cases of infinite and finite barrier where $\epsilon=\frac{\epsilon_{dot}}{\epsilon_{surrounding}}$.   (c) eigenvalues  including e-h coulomb and exchange interactions. The states are labelled with $F_{z}$ with superscripts standing for upper and lower state. The midpoint of the exciton energies with $M$=$\pm \frac{3}{2}$ and $M$=$\pm \frac{1}{2}$ is taken as the reference energy (zero of energy).
     
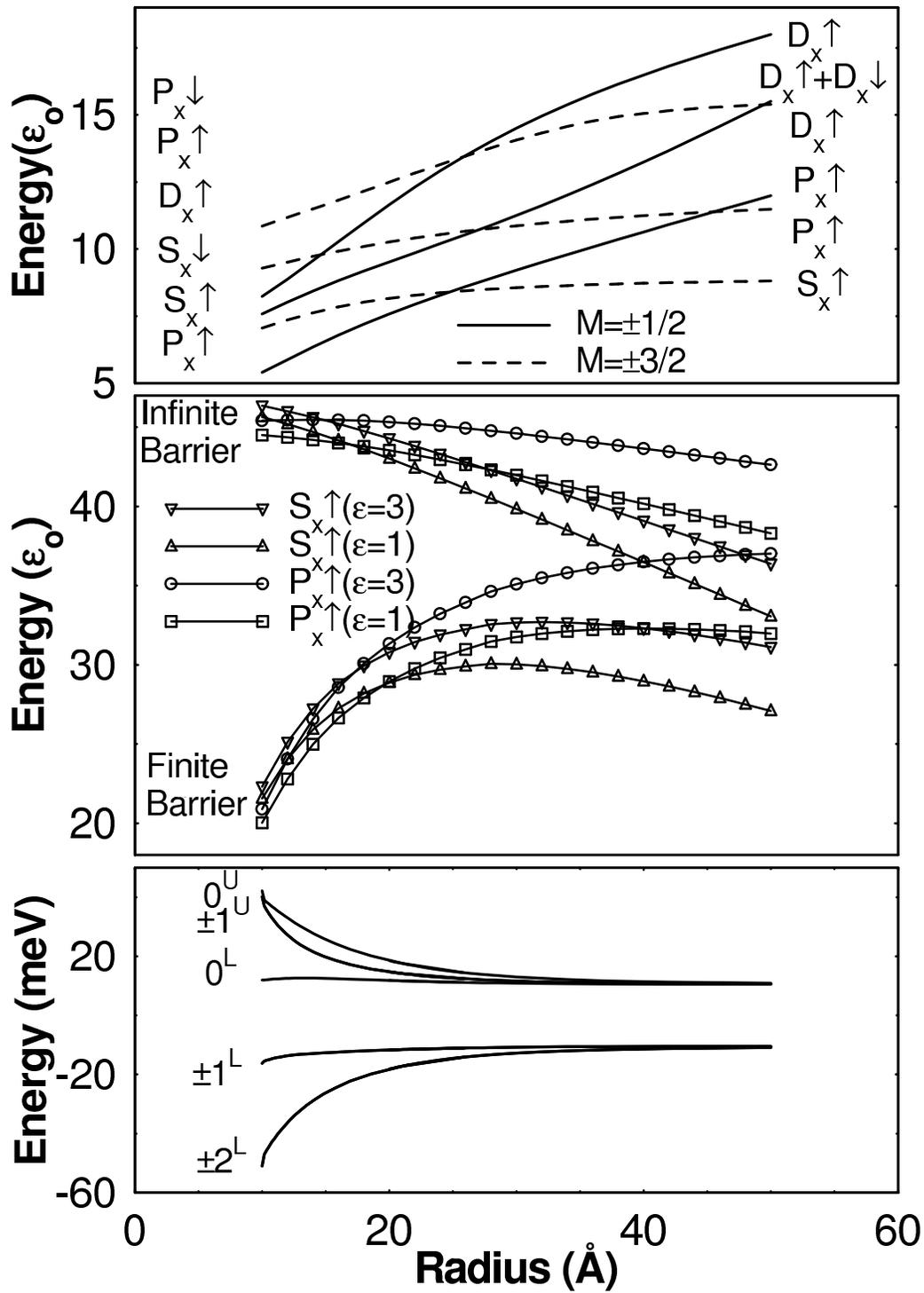
\noindent Fig.3:
The energy levels in units of $ \epsilon_{o} $ for wurtzite CdSe QDs as a function of dot radius R in the (a) absence and (b) presence of e-h coulomb interaction and dielectric mismatch effects(without exchange) for both the cases of infinite and finite barrier. The energy reference is taken as the top of the valence band with the spin orbit effects included. The states are labelled as $S_{x}\downarrow, P_{x}\uparrow$ etc, where capital letters represent  the dominant $\bf{L}$ present and the subscripts stand for X-, Y-like or Z-like states, the arrow indicating the spin state. (c) eigenvalues including the e-h coulomb and exchange interactions. 

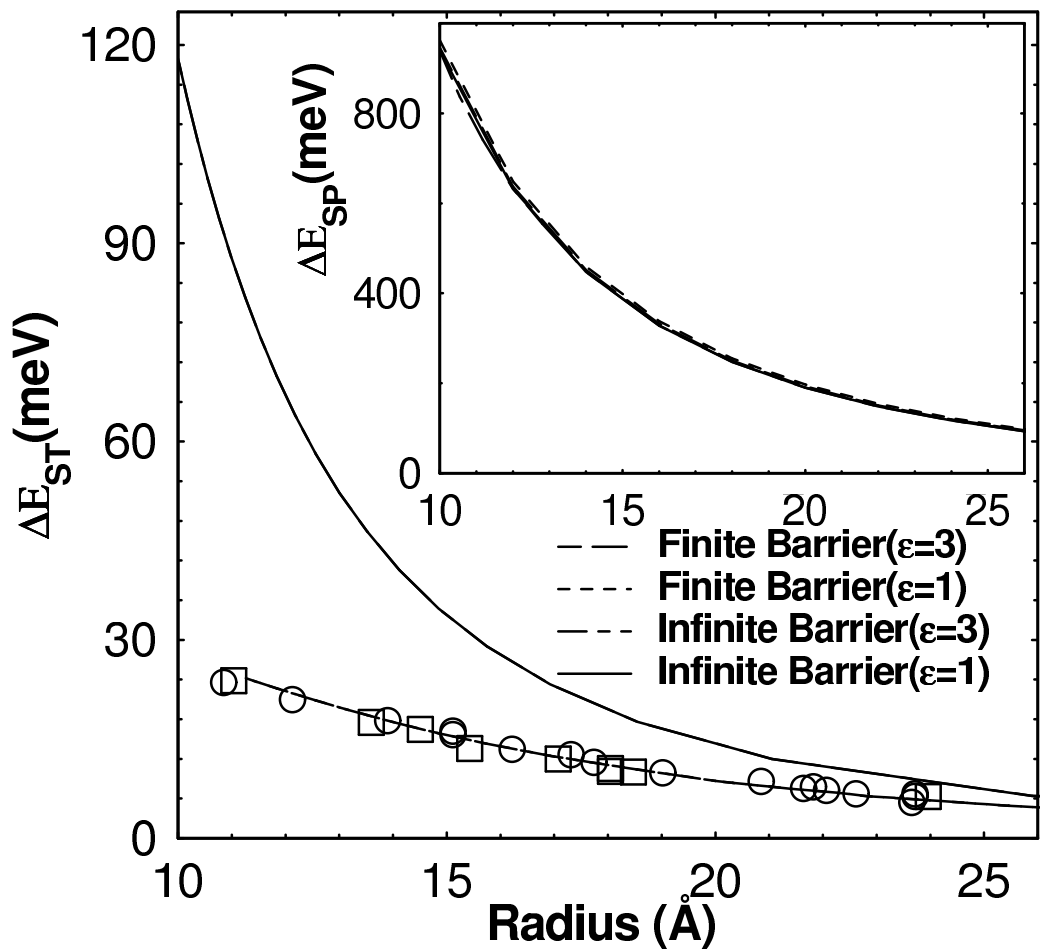
\noindent Fig.4:
Singlet triplet splitting in InAs in the absence $\epsilon=1$ and presence $\epsilon=3$ of dielectric mismatch effects for both the cases of infinite and finite barrier. The inset gives the energy differences of S and P states. Experimental data(empty circle and square) are taken from Ref.4. 

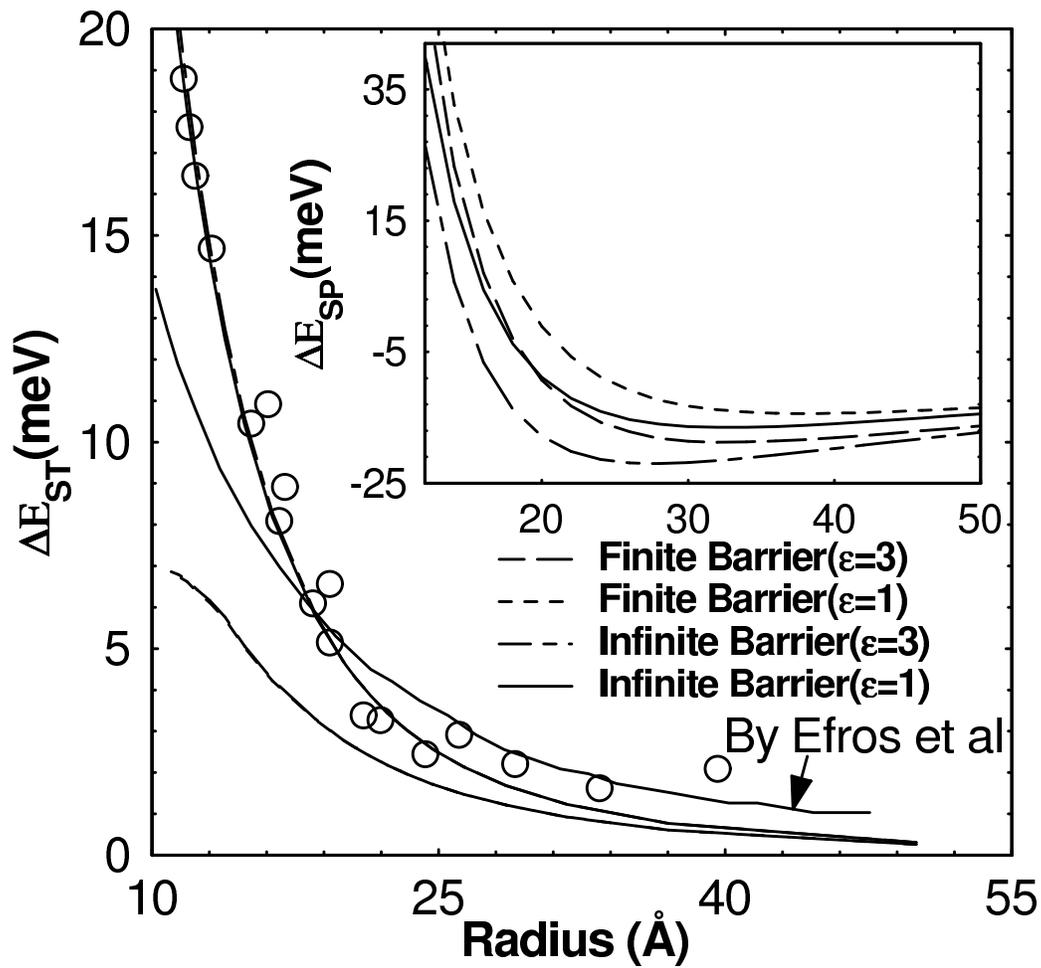
\noindent Fig.5:
The singlet triplet splitting in CdSe in the absence and presence of dielectric mismatch effects for both the cases of infinite and finite barrier. The inset gives the energy difference of S and P states with $M=\frac{3}{2}$ and $\frac{1}{2}$ respectively. Experimental data(empty circle) are taken from Ref.3.

\end{document}